\documentclass[12pt]{article}

\usepackage{graphicx}                     

\usepackage{xspace}                       

\newcommand{\ia}{{\"{\i}}}   
\newcommand{\absatz}{\vspace{2ex}\noindent}

\newcommand{\journal}[4]{{#1}\textbf{{#2}}, #3 (#4)}

\newcommand{\NPA}{Nucl.\ Phys.\ \textbf{A}}

\newcommand{\half}{\frac{1}{2}}

\newcommand{\ii}{\mathrm{i}}
\newcommand{\dd}{\mathrm{d}}
\newcommand{\tr}{\mathrm{tr}}
\newcommand{\T}{\mathrm{T}}

\newcommand{\pv}{\vec{\,\!p}\!\:{}}

\newcommand{\mpi}{m_\pi}
\newcommand{\fpi}{f_\pi}
\newcommand{\MeV}{\mathrm{MeV}}
\newcommand{\fm}{\mathrm{fm}}
\newcommand{\ENTNoPion}{ENT(${\pi\hskip-0.55em /}$)\xspace}
\newcommand{\NoPion}{{\pi\hskip-0.55em /}\xspace}
\newcommand{\upNoPion}{{}^{\pi\hskip-0.4em /}}
\newcommand{\NtwoLO}{N${}^2$LO\xspace}

\newcommand{\calA}{\mathcal{A}}
\newcommand{\calL}{\mathcal{L}}

\newcommand{\mytitle}[1]{
                         \begin{center}
                           \LARGE{\textbf{#1}}
                         \end{center}}
\newcommand{\myauthor}[1]{\textbf{#1}}
\newcommand{\myaddress}[1]{\textit{#1}}
\newcommand{\mypreprint}[1]{\begin{flushright} #1 \end{flushright}}

\begin{document}
%

\begin{titlepage}

  \mypreprint{
    nucl-th/0108060\\
    TUM-T39-01-21\\
    26th August 2001}

  \vspace*{1cm}

  \mytitle{An Introduction to Few Nucleon Systems\\
    in Effective Field Theory\thanks{Invited plenary talk at the Conference on
      Mesons and Light Nuclei 2001, Praha (Czech Republic) 2nd -- 6th July
      2001; to be published in the Proceedings; preprint numbers
      nucl-th/0108060, TUM-T39-01-21.}}
  
  \vspace*{0.5cm}

\begin{center}
  
  \myauthor{Harald W.\ Grie\3hammer}\footnote{Email:
    hgrie@physik.tu-muenchen.de}
  
  \vspace*{0.5cm}
  
  \myaddress{
    Institut f{\"u}r Theoretische Physik (T39), Physik-Department,\\
    Technische Universit{\"a}t M{\"u}nchen, D-85747 Garching, Germany}
  
  \vspace*{0.2cm}

\end{center}

\vspace*{0.5cm}

\begin{abstract}
  Progress in the Effective Field Theory of two and three nucleon systems is
  sketched, concentrating mainly on the low energy version in which pions are
  integrated out as explicit degrees of freedom. Examples given are: the
  extraction of nucleon polarisabilities from deuteron Compton scattering at
  very low energies; the energy dependence of the nucleon polarisabilities;
  three body forces and the triton; and $\mathrm{nd}$ partial waves at momenta
  below the pion cut.
\end{abstract}

\end{titlepage}

\setcounter{page}{2} \setcounter{footnote}{0} \newpage
  
%

\section{Foundations of Effective Nuclear Theory}

This presentation is a cartoon of the Effective Field Theory (EFT) of two and
three nucleon systems as it emerged in the last three years, using a lot of
words and figures, and a few cheats. For details, I refer to the bibliography
of a recent review~\cite{review}, and to papers with
Th.R.~Hemmert~\cite{polarisabilitiesrunning}, G.~Rupak~\cite{comptonnopions},
J.-W.~Chen, R.P.~Springer and M.J.~Savage~\cite{Compton},
P.F.~Bedaque~\cite{pbhg,pbfghg}, and F.~Gabbiani~\cite{pbfghg}. I will mainly
concentrate on the theory in which pions are integrated out as explicit
degrees of freedom, but also comment on the extensions to include pions. Of
the plenary talks covering related subjects, E.~Epelbaum's
contribution~\cite{Epelbaum} investigates Weinberg's proposal to include pions
in more detail for the few nucleon system, D.R.~Phillips~\cite{Phillips}
concentrates on electro-magnetic reactions on the deuteron,
R.~Timmermans~\cite{Timmermans} explains how chiral symmetry is seen in a
modern potential and phase shift analysis, and M.~Birse finally provides
background on the renormalisation group point of view~\cite{Birse}.

For want of free neutron targets, one cannot na{\ia}vely extract fundamental
iso-scalar and iso-vector properties of the nucleon separately from
experiment.  On the other hand, of all nuclei, the deuteron comes closest to
an iso-scalar target and hence appears well suited to extract nucleon and --
after removing proton effects -- neutron properties. Still, the analysis is
not straightforward because, however small the deuteron binding energy seems,
binding effects are often not negligible at low energies where properties of
the static nucleon are tested.  Although the impulse approximation of treating
the nucleons inside the deuteron as quasi-free is bound to become the better
the higher the typical momentum scale of the process is, the improved
resolution also necessitates a more detailed description of both the binding
between and structure of the nucleons: Effects from meson exchanges and
excited states are resolved at intermediate scales, and at even higher
energies, the nucleons and mesons themselves dissolve into quarks and gluons.

Nonetheless, model-independent predictions and extractions at low energies can
succeed because nuclear physics provides a separation of scales.  This
observation is a cornerstone of the Effective Field Theory approach.

Effective Field Theory methods are largely used in many branches of physics
where a separation of scales exists. In low energy nuclear
systems, the scales are, on one side, the low scales of the typical momentum
of the process considered and the pion mass, and on the other side the higher
scales associated with chiral symmetry and confinement. This separation of
scales produces a low energy expansion, resulting in a description of strongly
interacting particles which is systematic and rigorous. It is also model
independent (meaning, independent of assumptions about the non-perturbative
QCD dynamics): Given that QCD is the theory of strong interactions and that
chiral symmetry is broken via the Goldstone mechanism, Wilson's
renormalisation group arguments show that there is only one local low energy
field theory which originates from it: Chiral Perturbation Theory and its
extension to the many-nucleon system discussed here.

\subsection{The Lagrangean}

Three main ingredients enter the construction of an EFT: the Lagrangean, the
power counting and a regularisation scheme.  First, the relevant degrees of
freedom are identified. In his original suggestion how to extend EFT
methods to systems containing two or more nucleons, Weinberg~\cite{Weinberg}
noticed that below the $\Delta$ production scale, only nucleons and pions need
to be retained as the infrared relevant degrees of freedom of low energy QCD.
Because at these scales the momenta of the nucleons are small compared to
their rest mass, the theory becomes non-relativistic at leading order in the
velocity expansion, with relativistic corrections systematically included at
higher orders. The most general chirally invariant Lagrangean consists hence
of contact interactions between non-relativistic nucleons, and between
nucleons and pions, with the first terms reading
\begin{eqnarray}\label{ksw}
   {\mathcal{L}}_\mathrm{NN}&=&N^{\dagger}(\ii D_0+\frac{\vec{D}^2}{2M})N+
   \;\frac{\fpi^2}{8}\;
   \tr[(D_{\mu} \Sigma^{\dagger})( D^{\mu} \Sigma)]\;+
   \;g_A N^{\dagger} \vec{A}\cdot\vec{\sigma} N\;-
   \\
   &&-\;C_0 (N^{\T} P^i N)^{\dagger} \ (N^{\T} P^i N)\;
   + \;\frac{C_2}{8}
   \left[(N^{\T} P^i N)^{\dagger} (N^{\T} P^i
     (\stackrel{\scriptscriptstyle\rightarrow}{D}-
      \stackrel{\scriptscriptstyle\leftarrow}{D})^2 N)+
   \mathrm{H.c.}\right]
   + \dots,\nonumber
\end{eqnarray}
where $N={p\choose n}$ is the nucleon doublet of two-component spinors and
$P^i$ is the projector onto the iso-scalar-vector channel, $
P^{i,\,b\beta}_{a\alpha}=\frac{1}{\sqrt{8}}
(\sigma_2\sigma^i)_{\alpha}^{\beta} (\tau_2)_a^b$.  The iso-vector-scalar part
of the $\mathrm{NN}$ Lagrangean introduces more constants $C_i$ and
interactions and has not been displayed for convenience. The field
$\xi(x)=\sqrt{\Sigma}=e^{\ii \Pi/\fpi}$ describes the pion with a decay
constant $\fpi=130\;\MeV$.  $D_{\mu}=\partial_{\mu}+V_{\mu}$ is the chirally
covariant derivative, and
$A_{\mu}=\frac{\ii}{2}(\xi\partial_{\mu}\xi^{\dagger}-\xi^{\dagger}
\partial_{\mu}\xi)$
($V_\mu=\frac{1}{2}(\xi\partial_{\mu}\xi^{\dagger}+\xi^{\dagger}
\partial_{\mu}\xi)$) the axial (vector) pionic current. The interactions
involving pions are severely restricted by chiral invariance. As such, the
theory is an extension of Chiral Perturbation Theory and Heavy Baryon Chiral
Perturbation Theory to the many nucleon system: The terms in the first line
couple only pions amongst themselves, and to one nucleon, providing the
familiar nuclear long range force. The terms of the second line couple two
nucleons to each other and to pions via point-like interactions. Like in its
cousins, the coefficients of the low energy Lagrangean encode all short
distance physics -- branes and strings, quarks and gluons, resonances like the
$\Delta$ or $\rho$ -- as strengths of the point-like interactions between
particles. As it is not possible yet to derive these constants by solving QCD
e.g.~on the lattice, the most practical way to determine them is by fitting to
experiment.

\subsection{The Power Counting}

Because the Lagrangean (\ref{ksw}) consists of infinitely many terms
only restricted by symmetry, an EFT may at first sight suffer from lack of
predictive power. Indeed, as the second part of an EFT formulation, predictive
power is ensured by establishing a power counting scheme, i.e.\ a way to
determine at which order in a momentum expansion different contributions will
appear, and keeping only and all the terms up to a given order. The
dimensionless, small parameter on which the expansion is based is the typical
momentum $Q$ of the process in units of the scale $\Lambda_\mathrm{NN}$ at
which the theory is expected to break down.  Values for $\Lambda_\mathrm{NN}$
and $Q$ have to be determined from comparison to experiments and are a priori
unknown. Assuming that all contributions are of natural size, i.e.\ ordered by
powers of $Q$, the systematic power counting ensures that the sum of all terms
left out when calculating to a certain order in $Q$ is smaller than the last
order retained, allowing for an error estimate of the final result.

For extremely small momenta $Q\ll\mpi$, the pion does not enter as explicit
degree of freedom describing long range forces. All its effects are
absorbed into the coefficients $\upNoPion C_i$, while formally $\mpi\to\infty$
in (\ref{ksw}). The only forces between nucleons are thus point-like two and
more nucleon interactions with strengths $\upNoPion C_i$.  This Effective
Nuclear Theory with pions integrated out (\ENTNoPion,~\cite{CRS}) was recently
pushed to very high orders in the two-nucleon sector where accuracies of the
order of $1\%$ were obtained. It can be viewed as a systematisation of
Effective Range Theory with the inclusion of relativistic and short distance
effects traditionally left out in that approach. Because of non-analytic
contributions from the pion cut, the breakdown scale of this theory must be of
the order $\upNoPion\Lambda_\mathrm{NN}\sim\mpi$. For the ENT with explicit
pions, we would suspect the breakdown scale to be of the order of $M_\Delta-M$
or $m_\rho$, as $\Delta$ and $\rho$ are not explicit degrees of freedom
in (\ref{ksw}).

Even if calculations of nuclear properties were possible starting from the
underlying QCD Lagrangean, EFT simplifies the problem considerably by
factorising it into a long distance part which contains the infrared-relevant
physics and is dealt with by EFT methods and a short distance part, subsumed
into the coefficients of the Lagrangean. QCD therefore ``only'' has to provide
these constants, avoiding full-scale calculations of e.g.~bound state
properties of two nuclear systems using quarks and gluons. EFT provides an
answer of finite accuracy because higher order corrections are systematically
calculable and suppressed in powers of $Q$. Hence, the power counting allows
for an error estimate of the final result, with the natural size of all
neglected terms known to be of higher order. Relativistic effects, chiral
dynamics and external currents are included systematically, and extensions to
include e.g.~parity violating effects are straightforward.  Gauged
interactions and exchange currents are unambiguous.  Results obtained with EFT
are easily dissected for the relative importance of the various terms.
Because only $S$-matrix elements between on-shell states are observables,
ambiguities nesting in ``off-shell effects'' are absent. On the other hand,
because only symmetry considerations enter the construction of the Lagrangean,
EFTs are less restrictive as no assumption about the underlying QCD dynamics
is incorporated. Hence the proverbial quib that ``EFT parameterises our
ignorance''.

\absatz In systems involving two or more nucleons, establishing a power
counting is complicated because unnaturally large scales have to be
accommodated: Given that the typical low energy scale in the problem should be
the mass of the pion as the lightest particle emerging from QCD, fine tuning
is required to produce the large scattering lengths in the $\mathrm{S}$ wave
channels ($1/a^{{}^1\mathrm{S}_0}=-8.3\;\MeV,\;
1/a^{{}^3\mathrm{S}_1}=36\;\MeV$).  Since there is a bound state in the
${}^3\mathrm{S}_1$ channel with a binding energy $B=2.225\;\MeV$ and hence a
typical binding momentum $\gamma=\sqrt{M B}\simeq 46\;\MeV$ well below the
scale $\Lambda_\mathrm{NN}$ at which the theory should break down, it is also
clear that at least some processes have to be treated non-perturbatively in
order to accommodate the deuteron, i.e.~an infinite number of diagrams has to
be summed or equivalently, a Schr\"odinger equation needs to be solved.

\begin{figure}[!htb]
    \centerline{\includegraphics*[width=0.9\textwidth]{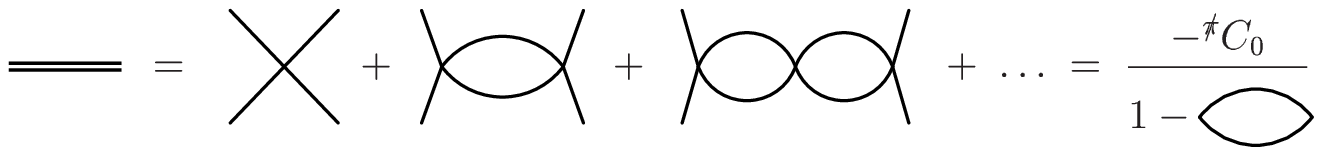}}
    \caption{Re-summation of the contact interactions into the deuteron
      propagator.}
    \label{fig:deuteronprop}
\end{figure}

For simplicity, let us first turn to \ENTNoPion in which the pion is
integrated out. Here, a way to incorporate this fine tuning into the power
counting was suggested by Kaplan, Savage and Wise~\cite{KSW}, and by van
Kolck~\cite{BiraAleph}. At very low momenta, contact interactions with several
derivatives -- like $p^2\,\upNoPion C_2$ -- should become unimportant, and we
are left only with the contact interactions proportional to $\upNoPion C_0$.
The dominating contribution to nucleons scattering in an $\mathrm{S}$ wave
comes hence from two nucleon contact interactions and is summed geometrically
in Fig.~\ref{fig:deuteronprop} in
order to produce the shallow real bound state.

How to justify this? Dimensional analysis allows the size of any diagram to be
estimated by scaling momenta by a factor of $Q$ and non-relativistic kinetic
energies by a factor of $Q^2/M$.  The remaining integral includes no
dimensions and is taken to be of the order $Q^0$ and of natural size. This
scaling implies the rule that nucleon propagators contribute one power of
$M/Q^2$ and each loop a power of $Q^5/M$.  Assuming that
\begin{eqnarray}\label{scalingksw}
  \upNoPion C_0\sim\frac{1}{M Q}&\;\;,\;\;&
  \upNoPion C_2\sim\frac{1}{M \;\upNoPion\Lambda_\mathrm{NN} Q^2}\;\;,
\end{eqnarray}
the diagrams contributing at leading order to the deuteron propagator are
indeed an infinite number as shown in Fig.~\ref{fig:deuteronprop}, each one of
order $1/(MQ)$. The deuteron propagator
\begin{equation}
  \label{deuteronpropagator}
  \frac{4\pi}{M}\;\frac{-\ii}{\frac{4\pi}{M\,\upNoPion C_0}+\mu-
    \sqrt{\frac{\pv^2}{4}-Mp_0-\ii\varepsilon}}
\end{equation}
has the correct pole position and cut structure when one chooses
$
\upNoPion C_0(\mu)=\frac{4\pi}{M}\;\frac{1}{\gamma-\mu}
$.

$\upNoPion C_0$ becomes dependent on an arbitrary scale $\mu$ because of the
regulator dependent, linear UV divergence in each of the bubble diagrams in
Fig.~\ref{fig:deuteronprop}.  Indeed, when choosing $\mu\sim Q$, the leading
order contact interaction scales as in (\ref{scalingksw}). As
expected for a physical observable, the $\mathrm{NN}$ scattering amplitude
becomes independent of $\mu$, the renormalisation scale or cut-off chosen.
All other coefficients $\upNoPion C_i$ can be shown to be higher order, so
that the scheme is self-consistent. Observables are independent of the cut-off
chosen.

The linear divergence of each bubble in Fig.~\ref{fig:deuteronprop} does not
show in dimensional regularisation as a pole in $4$ dimensions, but it does
appear as a pole in $3$ dimensions which we subtract following the Power
Divergence Subtraction scheme~\cite{KSW}.  Dimensional regularisation is
chosen to explicitly preserve the systematic power counting as well as all
symmetries (esp.~chiral invariance) at each order in every step of the
calculation. At leading (LO), next-to-leading order (NLO) and often even
N${}^3$LO in the two nucleon system, it also allows for simple, closed answers
whose analytic structure is readily asserted. Power Divergence Subtraction
moves hence a somewhat arbitrary amount of the short distance contributions
from loops to counterterms and makes precise cancellations manifest which
arise from fine tuning~\cite{BiraAleph}.

\subsection{Including Pions}

The power counting of the zero and one nucleon sector of the theory are fixed
by Chiral Perturbation Theory and its extension to the one baryon sector.
Therefore, the question to be posed is: How does the power counting of the
contact terms $\upNoPion C_i$ change above the pion cut, i.e.~when the pion
must be included as explicit degree of freedom?

\begin{figure}[!ht]
  \centerline{\includegraphics*[width=0.5\linewidth]{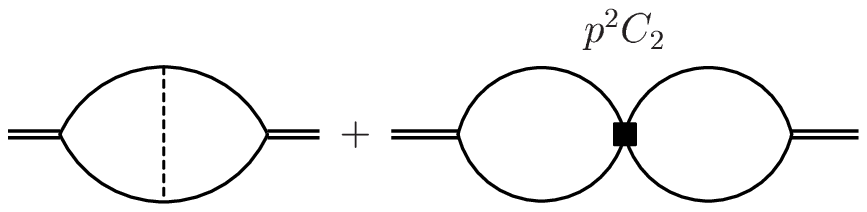}}
          \caption{The next-to-leading order in ENT(KSW).}
    \label{fig:NLOKSW}
\end{figure}

If pulling out pion effects does not affect the running of $C_0$ too much, one
surprising result arises: Because chiral symmetry implies a derivative
coupling of the pion to the nucleon at leading order, the instantaneous one
pion exchange scales as $Q^0$ and is {\it smaller} than the contact piece
${}^\mathrm{KSW}C_0\sim \upNoPion C_0 \sim Q^{-1}$. Pion exchange and higher
derivative contact terms appear hence only as perturbations at higher orders.
The LO contribution in this scheme is still given by the geometric series in
Fig.~\ref{fig:deuteronprop}. In contradistinction to iterative potential model
approaches, each higher order contribution is inserted only once. In this
scheme, the only non-perturbative physics responsible for nuclear binding is
extremely simple, and the more complicated pion contributions are at each
order given by a finite number of diagrams. For example, the NLO contributions
to $\mathrm{NN}$ scattering in the ${}^1\mathrm{S}_0$ channel are the one
instantaneous pion exchange and the two nucleon interaction with two
derivatives, Fig.~\ref{fig:NLOKSW}.  The constants are determined e.g.~by
demanding the correct scattering length and effective range. This approach is
known as the ``KSW'' counting of ENT, ENT(KSW)~\cite{KSW}.
    
\begin{figure}[!ht]
  \centerline{\includegraphics*[width=0.9\linewidth]{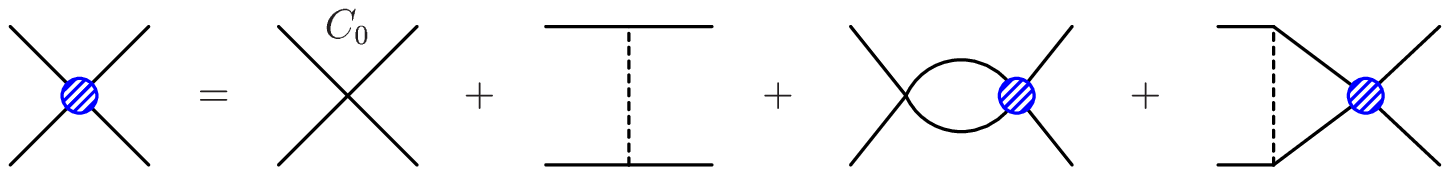}}
    \caption{The leading order in ENT(Weinberg).}
    \label{fig:weinberg}
\end{figure}

If on the other hand, the power counting for $\upNoPion C_0$ is dramatically
modified when one includes pions, both the $C_0$ interactions and the one pion
exchange might have to be iterated. In his paper, Weinberg~\cite{Weinberg}
therefore suggested to power count not the amplitude -- as is usually done in
EFT -- but the potential, and then to solve a Schr\"odinger equation with a
chiral potential, as pictorially represented in Fig.~\ref{fig:weinberg}.
E.~Epelbaum's talk~\cite{Epelbaum} gives more details on the results obtained
so far in this approach.

Both power countings are presently under investigation for consistency and
convergence, and each has its advantages and shortcomings~\cite{review}.
Recently, Beane et al.~\cite{Beane:2001bc} demonstrated that Weinberg's power
counting is not self-consistent in the spin doublet channel of $\mathrm{NN}$
scattering and that pions are perturbative there, although convergence is
slow. On the other hand, the strongly attractive $1/r^3$ part of the pionic
tensor force in the spin triplet channel necessitates a non-perturbative
renormalisation of the one pion exchange. In that channel, the KSW counting is
hence inconsistent, and Weinberg's proposal is chosen by Nature. It is not yet
clear how this observation is to be extended to the power counting of counter
terms in which two or more nucleons couple on external currents. After all,
Nuclear Physics in the two body system is more than $\mathrm{NN}$ scattering.

Although in general process dependent, the expansion parameter is found to be
of the order of $\frac{1}{3}$ in both approaches and in most applications, so
that NLO calculations can be expected to be accurate to about $10\%$, and \NtwoLO
calculations to about $4\%$.  In all cases, experimental agreement is within
the estimated theoretical uncertainties, and in some cases, previously unknown
counterterms could be determined.

\section{Applications}

\subsection{Polarisabilities from Low Energy Deuteron Compton Scattering}

The first example we turn to demonstrates not only how simple it is to compute
processes involving both external gauge and exchange currents in a
self-consistent way, but also how ``effective'' the power counting is to
estimate theoretical uncertainties. The following calculation is
model-independent and performed in \ENTNoPion~\cite{comptonnopions}.

\begin{figure}[!htb]
  \centerline{
    \parbox[b][0.3\textwidth][t]{0.39\textwidth}{
    \includegraphics*[width=0.9\linewidth]{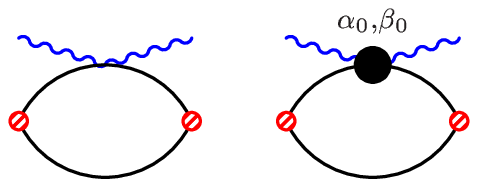}
    }
  \hfill
    \includegraphics*[width=0.55\textwidth]{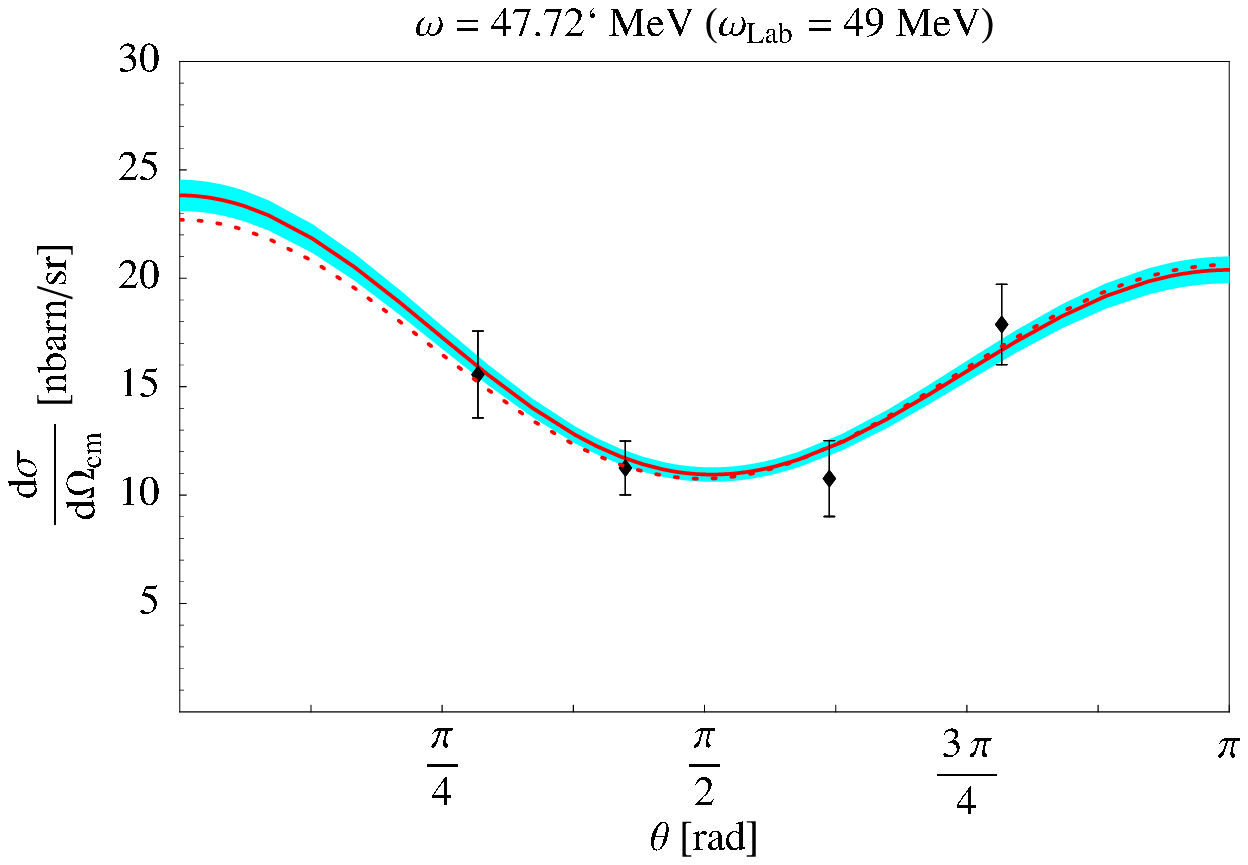}}
\caption{Left: Thomson term and polarisabilities contribution in deuteron
  Compton
  scattering. Right: \ENTNoPion result for the differential cross section,
  fitted to data. Dashed: \protect$\alpha_0$ and \protect$\beta_0$ fitted
  independently; solid: \protect$\alpha_0$ and \protect$\beta_0$ constrained
  by the Baldin sum rule. The estimate of the accuracy of the calculation at
  \NtwoLO (\protect$\pm 3\%$) is indicated by the shaded area.}
\label{fig:compton}
\end{figure}

The iso-scalar, scalar electric and magnetic dipole nucleon polarisabilities
parameterise the deformation of the nucleon in an external electro-magnetic
dipole field:
\begin{equation}
  \label{polarisabilitiesdef}
  \calL_\mathrm{pol}=2\pi\;(\alpha_0 \vec{E}^2 +\beta_0\vec{B}^2)
  \;N^\dagger N\;\;,
\end{equation}
where $\alpha_0:=(\alpha^{(p)}+\alpha^{(n)})/2$ and
$\beta_0:=(\beta^{(p)}+\beta^{(n)})/2$. How accurate does one have to measure
elastic Compton scattering of photons on the deuteron to see them, and how
severe are the binding effects which distinguish the deuteron from an
iso-scalar target? Compare the graphs containing these contact interactions of
\emph{a priori} unknown strengths to the Thomson term which constitutes LO in
accordance with the low energy theorem. A quick look on the left hand side of
Fig.~\ref{fig:compton} reveals that the former is suppressed against the
latter by a factor $\omega^2/\upNoPion\Lambda_{\mathrm{NN}}^2$, where $\omega$
is the photon energy and the breakdown scale of \ENTNoPion enters to get a
dimensionless ratio. Thus, the lower the photon energy $\omega$ in Compton
scattering, the less do $\alpha$ and $\beta$ contribute. On the other hand, at
$\omega\sim\gamma$, the polarisabilities contribute about
$\frac{1}{3}^2\sim10\%$ to the Compton cross section, while higher energies
introduce large theoretical errors since corrections from non-analytic pion
contributions are not suppressed sufficiently strong any more.

Results of an analytic calculation of the differential cross section to
\NtwoLO, i.e.~to an accuracy of $\sim 3\%$, at photon energies $\omega$ in the
window of opportunity at $15-50\;\MeV$ are presented in
Fig.~\ref{fig:compton}. The polarisabilities $\alpha_0$ and $\beta_0$ enter as
the only free parameters and carry a theoretical uncertainty of about $20\%$.
As a feasibility study, we used data at $\omega_\mathrm{Lab}=49\;\MeV$ to find
$\alpha_0=8.4\pm 3.0(\mathrm{exp})\pm 1.7(\mathrm{theor})$, $\beta_0=8.9\pm
3.9(\mathrm{exp})\pm 1.8(\mathrm{theor})$, each in units of $10^{-4}\;\fm^3$.
With the experimental constraint for the iso-scalar Baldin sum rule ($\alpha_0
+ \beta_0=14$), $\alpha_0=7.2\pm 2.1(\mathrm{exp})\pm 1.6(\mathrm{theor})$,
$\beta_0=6.9\mp 2.1(\mathrm{exp})\mp 1.6(\mathrm{theor})$. These values differ
from the proton ones, $\alpha^{(p)}=12\approx 10\;\beta^{(p)}$. The error bars
from the experimental uncertainty in our extraction are large; furthermore,
several conflicting measurements exist for the nucleon (and neutron)
polarisabilities, see~\cite{comptonnopions} for details. A more accurate
result can be achieved (i) predominantly by better data, and (ii)
by a higher order theoretical calculation including contributions from so far
undetermined two-nucleon-two-photon operators.  Although the scarcity of data
at low energies is a big hindrance, the comparison of the calculation to
experiment shows that it is quite appropriate to determine scalar
polarisabilities from very low energy Compton scattering.  The energy r\'egime
proposed is hence an interesting window of opportunity to determine nucleon
polarisabilities in a model-independent way without having to deal with pions
as explicit degrees of freedom.

Compton scattering has also been investigated in
ENT(KSW)~\cite{Compton} and in ENT(Weinberg), see
also~\cite{Phillips}. Surprisingly, the results all agree within the
theoretical uncertainty of the calculations even at relatively high momenta.
This seems to suggest that non-analytic pion contributions from meson exchange
diagrams are small. However, the higher the photon energy, the less seem the
polarisabilities extracted to be in agreement with the numbers obtained at
very low photon energy, see e.g.~\cite{Phillips}.

\subsection{Dynamical Polarisabilities and Compton Scattering}

\begin{figure}[!htb]
  \centerline{
    \includegraphics*[width=0.45\textwidth]{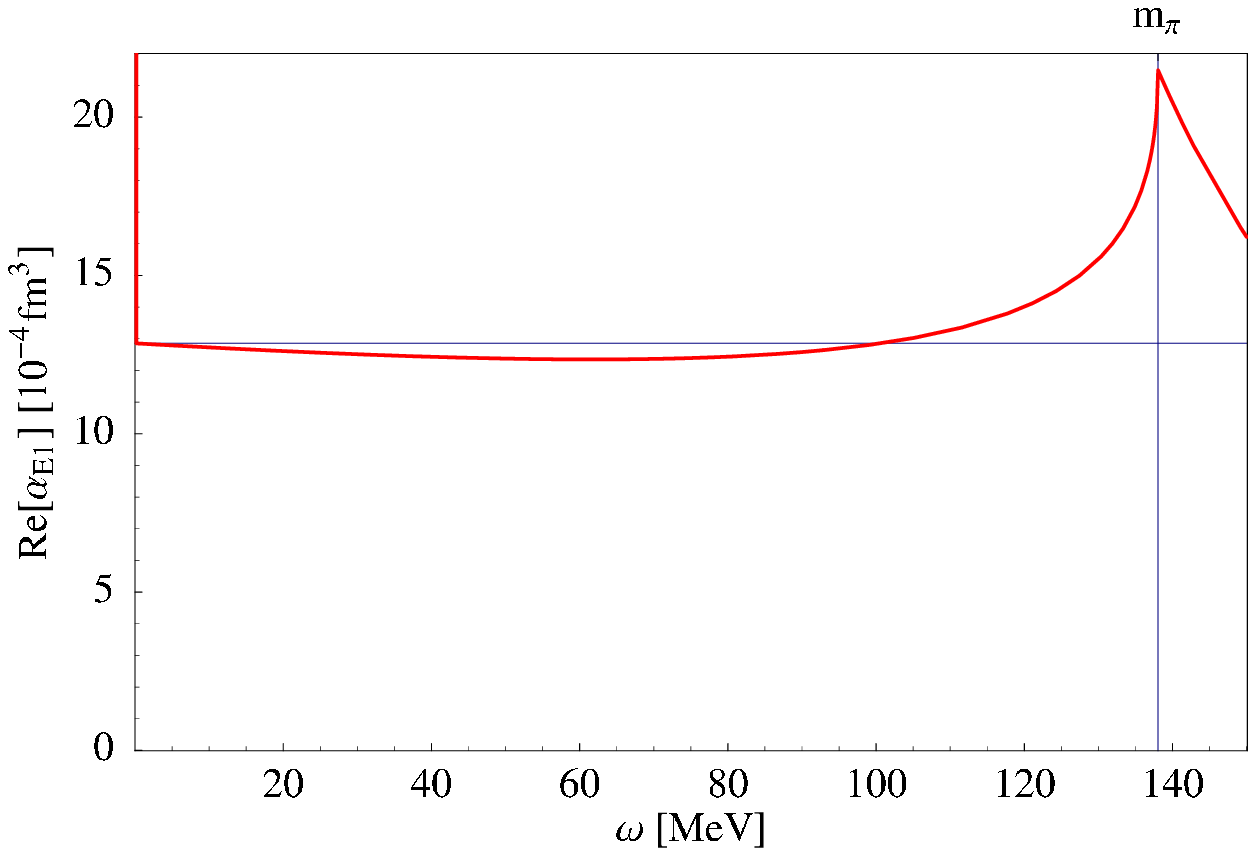}
    \hfill \includegraphics*[width=0.45\textwidth]{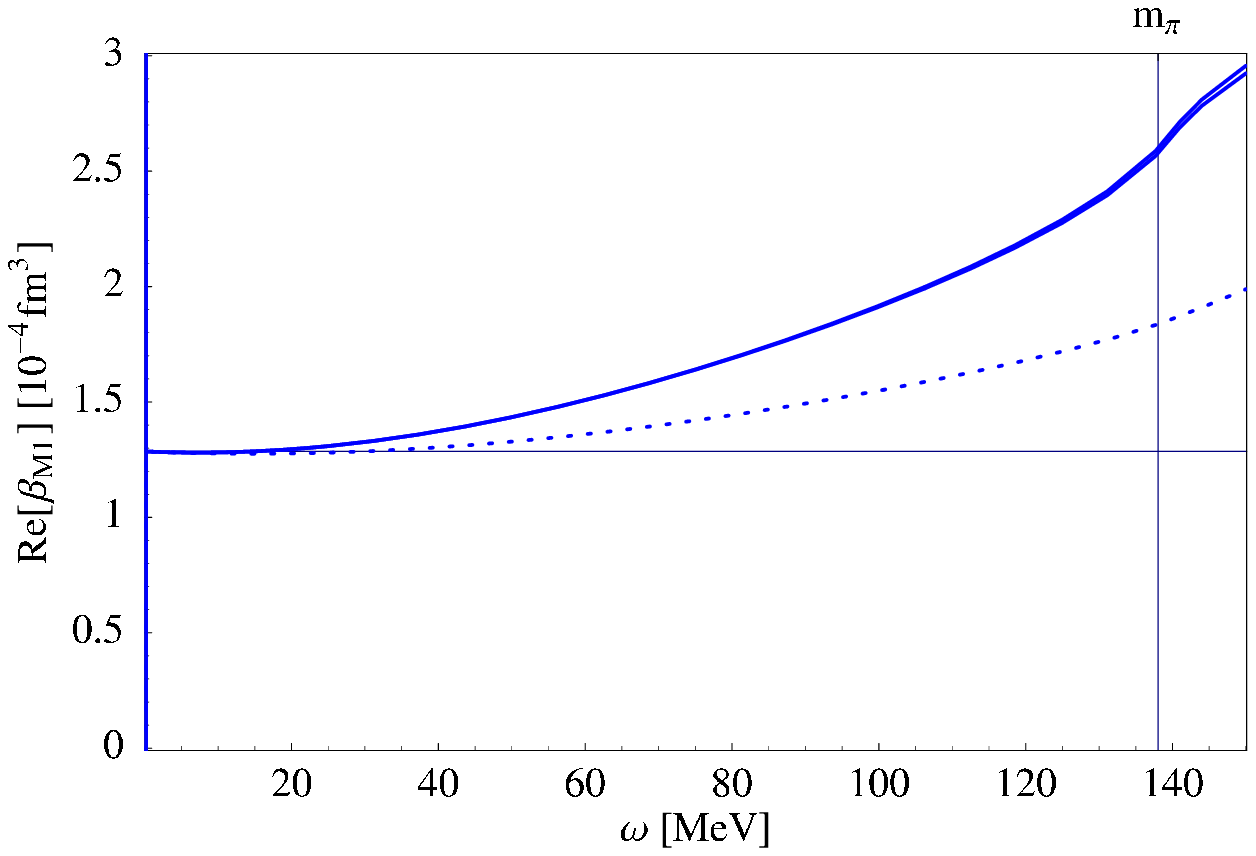}}
\caption{Leading order HB$\chi$PT predictions of the dependence of the
  dynamical electric and magnetic dipole polarisabilities on the photon
  energy. While there is no visible dependence of $\alpha_{E1}(\omega)$ on the
  number of multipoles included (left figure), the prediction for
  $\beta_{M1}(\omega)$ changes drastically when the extraction is truncated at
  the dipole polarisabilities (doted line in right figure), while including
  octupole polarisabilities makes no effect.}
\label{fig:polarisabilitiesrunning}
\end{figure}

One has to keep in mind~\cite{polarisabilitiesrunning} the well known fact
that polarisabilities themselves are energy dependent, probing the temporal
response of the system (so-called ``dynamical polarisabilities''). One expects
that the polarisabilities are enhanced around the pion production threshold
$\omega\sim\mpi$ since on-shell pions can then be produced out of the virtual
pion cloud around the nucleon. In addition, a resonance is expected at
$\omega\sim300\;\MeV$ because of contributions from the $\Delta(1232)$ as
resonance state of the nucleon. At very low energies, these effects contribute
corrections of order $\omega^2/(\mpi,\,M_\Delta-M_N)^2$, but as
$\omega\to\mpi$, they are expected to be large. For the proton and nucleon
magnetic dipole polarisability, they will be especially pronounced because it
is anomalously small compared to $\alpha^{(p)}$, and because the cancellation
of dia-magnetic contributions from the pion tail with para-magnetic
contributions from the $\Delta$ and pion core which is observed at zero energy
for the proton does not have to hold for the neutron, nor at finite energies.
As a first estimate, Fig.~\ref{fig:polarisabilitiesrunning} presents the
predictions of LO Heavy Baryon Chiral Perturbation Theory (no dynamical
$\Delta$s) and also estimates the effect of neglecting quadrupole and octupole
polarisabilities in the extraction. The effect on $\beta$ is sizeable. Thus, a
series of Compton scattering experiments at energies up to $\omega\sim\mpi$
can give valuable information on the competing dia- and para-magnetic effects
inside the nucleon.

\subsection{Three Body Forces and the Triton}

Since all interactions permitted by the symmetries must be included into the
Lagrangean, EFT dictates that in the three body sector, interactions like
\begin{equation}
  \label{threebodyterm}
  \calL_\mathrm{3body}=H\;(N^\dagger N)^3
\end{equation}
with unknown strength $H$ are present. Turning again to \ENTNoPion, we
therefore have to ask at which order in the power counting they will start to
contribute. Bedaque, Hammer and van Kolck~\cite{Stooges2} found the surprising
result that an unusual renormalisation makes the three body force of leading
order in the triton channel.

\begin{figure}[!htb]
    \centerline{\includegraphics*[width=0.7\textwidth]{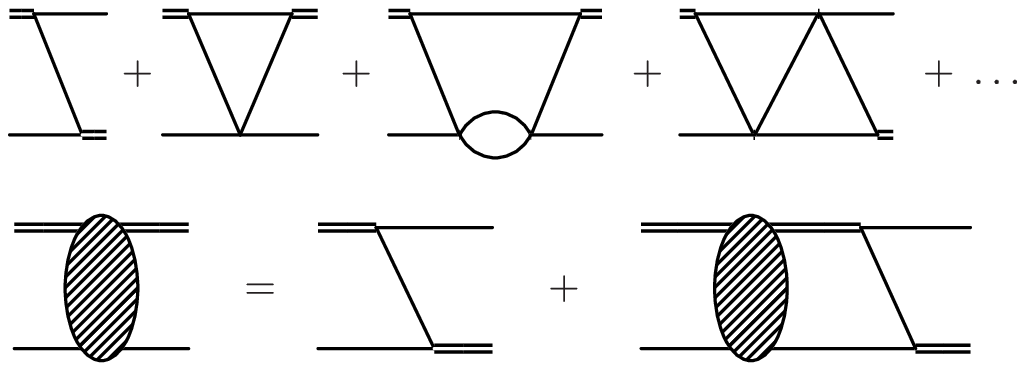}}
    \caption{The double infinite series of LO
      ``pinball'' diagrams, some of which are shown in the first line, is
      equivalent to the solution of the Faddeev equation of the second line.}
    \label{fig:LOfaddeev}
\end{figure}

In order to understand this finding, let us first consider the LO diagrams
which come from two nucleon interactions. The absence of Coulomb interactions
in the $\mathrm{nd}$ system ensures that only properties of the strong
interactions are probed.  All graphs involving only $\upNoPion C_0$
interactions are of the same order and form a double series which cannot be
written down in closed form.  Summing all ``bubble-chain'' sub-graphs into the
deuteron propagator, one can however obtain the solution numerically from the
integral equation pictorially shown in Fig.~\ref{fig:LOfaddeev} within seconds
on a personal computer.

Because one nucleon is exchanged in the intermediate state, each diagram is of
the order of the nucleon propagator, i.e.~$Q^{-2}$, while the first three
buddy force (\ref{threebodyterm}) seems to be of order $Q^0$. We are therefore
tempted to assume that three body forces are at worst \NtwoLO, i.e.~of the
order $10\%$. In the quartet channel, power counting suggests even N${}^4$LO
because the Pauli principle forbids three body forces without derivatives.
However, such na{\ia}ve counting was already fallacious in the two body sector
because of the presence of a low lying bound state and of a linear divergence
in each of the bubble graphs making up Fig.~\ref{fig:deuteronprop}. We
therefore investigate more carefully the UV behaviour of the amplitude
$\calA(k,p)$ at half off-shell momenta $p\gg k,\gamma$. The homogeneous part
of the integral equation simplifies then to
\begin{eqnarray}
  \label{UVfaddeev}
        \calA_{(l,s)}(0,p)&=&
        \frac{4\;\lambda(s)}{\sqrt{3}\,\pi}
          \int\limits_0^\infty \frac{\dd q}{p}
          \;\calA_{(l,s)}(0,q)\;
          Q_{l}\left[\frac{p}{q}+ \frac{q}{p}\right]\;\;,
\end{eqnarray}
where $\lambda(s)=-\frac{1}{2}$ in the spin quartet channel, and $1$ in
the doublet. $Q_l$ is the Legendre Polynomial of the second kind, $l$ the
angular momentum of the partial wave investigated.

\begin{figure}[!ht]
    \centerline{\includegraphics*[width=0.47\linewidth]{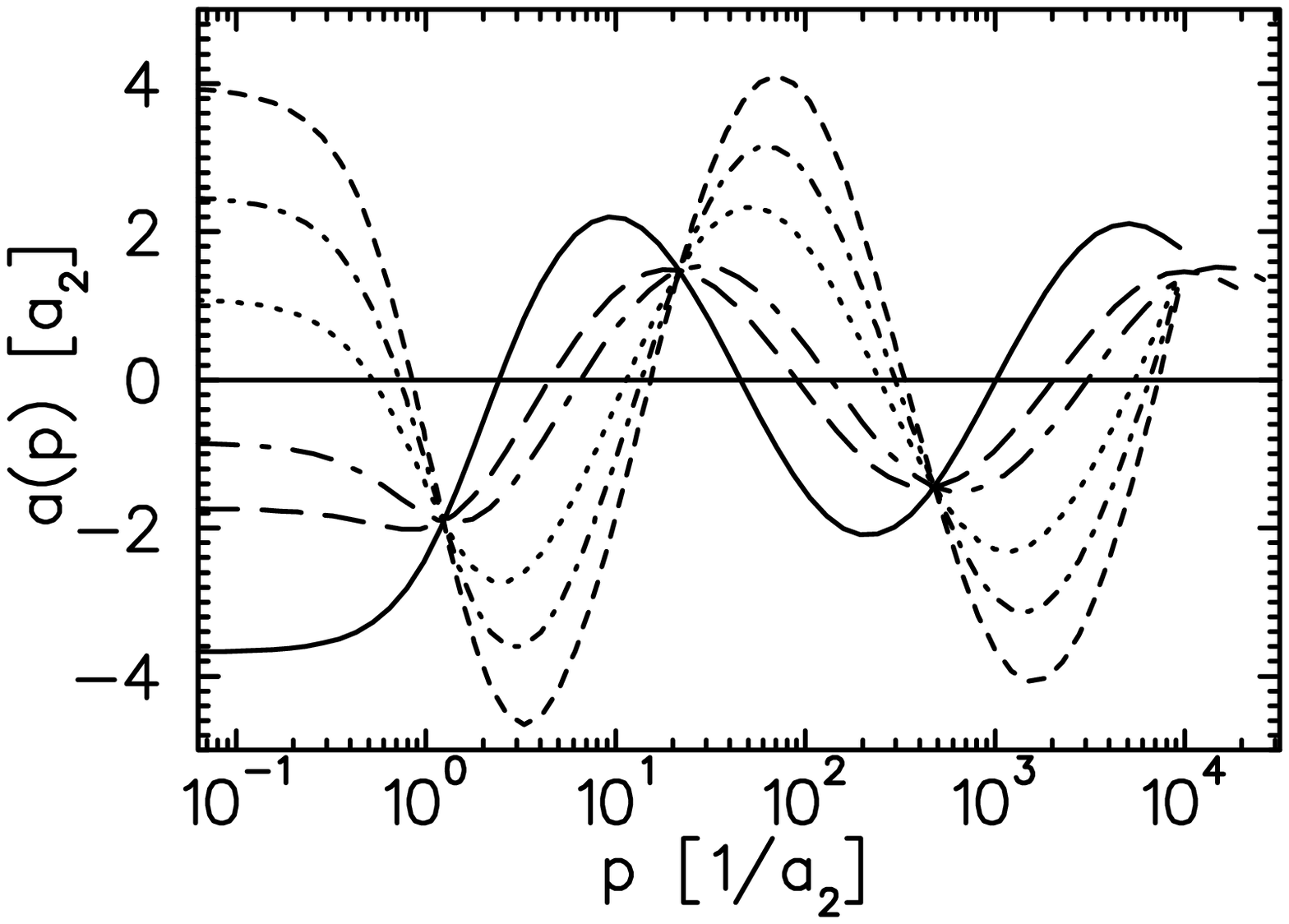}\hfill
    \includegraphics*[width=0.48\linewidth]{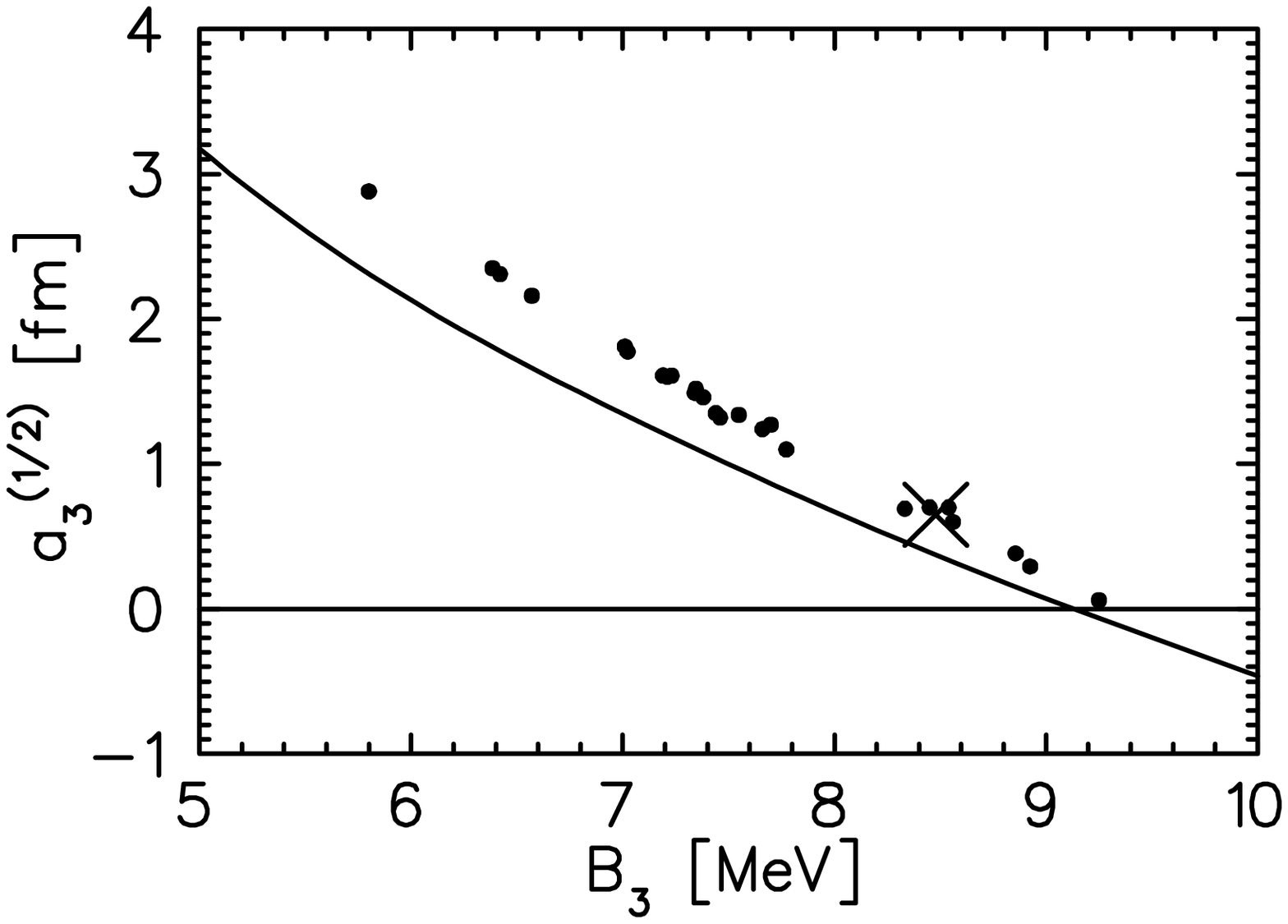}}
    \caption{Left: Variation of the cut-off by a factor 3 in a numerical study
      of the na{\ia}ve off-shell amplitude Fig.~\protect\ref{fig:LOfaddeev} in
      the triton channel changes the scattering length $a(p=0)$ dramatically.
      Right: Comparing the ENT prediction for the Phillips line with results
      from various potential models. Figures from
      Ref.~\protect\cite{Stooges2}.}
    \label{fig:Phillipsline}
\end{figure}

The equation is easily solved by a Mellin transformation,
$\calA_{(l,s)}(0,p)=p^{-1+s_0(l,s)}$, and indeed in most channels $s_0$ is real
so that only one solution exists which vanishes at infinite momentum and hence
is cut-off independent. However, the fact that the kernel of (\ref{UVfaddeev})
is not compact makes one suspicious whether this is always the case. Indeed,
there exists one and only one partial wave in which two linearly independent
solutions are found: the doublet $\mathrm{S}$ wave (triton) channel, where
$s_0(l=0,s=\half)=\pm1.0062\dots\ii$. Therefore in this channel, any
superposition with an arbitrary phase $\delta$ is also a solution,
\begin{equation}
  \label{UVsolution}
  \calA_{l=0,s=\half}(k\to 0,p)\propto\frac{\cos[1.0062 \ln p+\delta]}{p}\;\;.
\end{equation}
As each value of the phase provides a different boundary condition for the
solution of the full integral equation Fig.~\ref{fig:LOfaddeev}, the on-shell
amplitude $\calA(k,p=k)$ depends crucially on $\delta$. This means that the
on-shell amplitude seems sensitive to off-shell physics, and what is more, to
a phase which stems from arbitrarily high momenta. A numerical study of the
full half off-shell amplitude confirms these findings,
Fig.~\ref{fig:Phillipsline}. This cannot be.

\begin{figure}[!ht]
    \centerline{\includegraphics*[width=0.9\textwidth]{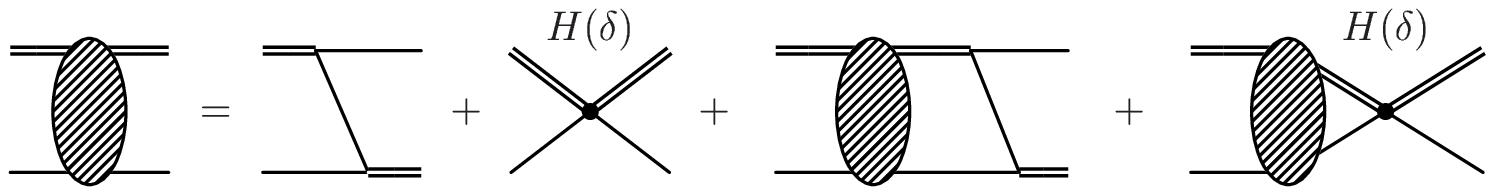}}
    \caption{The cut-off independent Faddeev equation in the
      triton channel.} 
    \label{fig:LOtriton}
\end{figure}

Since physics must be independent of the cut-off chosen, this sensitivity of
the on-shell amplitude on UV properties of the solution to
(\ref{fig:LOfaddeev}) must be remedied by adding a counter term. And since the
power counting in the two nucleon sector is fixed, 
a necessary and sufficient condition to render cut-off independent results is
to promote the three body force (\ref{threebodyterm}) to LO, $H(\delta)\sim
Q^{-2}$, and absorb all phase dependence into it, see Fig.~\ref{fig:LOtriton}.
How $H(\delta)$ varies with the cut-off or phase is known analytically
from (\ref{UVsolution}), but its initial value is unknown. Therefore, one
physical scale $\bar{\delta}$ must be determined experimentally. This one, new
free parameter explains why potential models which provide an accurate
description of $\mathrm{NN}$ scattering can vary significantly in their
predictions of the triton binding energy $B_3$ and three body scattering
length $a_3^{(1/2)}$ in the triton channel, although all of them lie on a
curve in the $(B_3,a_3^{(1/2)})$ plane, known as the Phillips line,
Fig.~\ref{fig:Phillipsline}. Determining $H(\bar{\delta})$ by fixing the three
body scattering length to its physical value, the triton binding energy is
found in \ENTNoPion to be $8.0\;\MeV$ at LO, $8.8\;\MeV$ at NLO, and the phase
shift in the triton channel is well in agreement with
experiment~\cite{HammerMehen}.

It must again be stressed that the three body force of strength $H(\delta)$
was added not out of phenomenological needs.  It cures the arbitrariness in
the off-shell and UV behaviour of the two body interactions which would
otherwise contaminate the on-shell amplitude.  Just as the off-shell behaviour
of the two body amplitude Fig.~\ref{fig:LOfaddeev}, the strength $H(\delta)$
is arbitrary, and only the sum of two and three body graphs is physically
meaningful. 

Summing the deuteron bubbles, each graph in the upper line of
Fig.~\ref{fig:LOfaddeev} behaves like $p^{-2}$ in the UV. But the solution of
the Faddeev equation goes like $p^{-1+s_0}$ with irrational (or even complex)
$s_0(l,s)$~\cite{hgtriton} and is hence more than the na{\ia}ve sum of graphs.

The limit cycle thus encountered in the triton channel is a new
renormalisation group phenomenon and also explains the Efimov and Thomas
effects~\cite{review,Stooges2}. In all other partial waves, three body forces
enter only at higher orders~\cite{hgtriton}. In contradistinction to the
explanation given above, J.~Gegelia's view on the triton is that a unique
solution can be constructed using involved numerical methods, and that the
three body force is not LO~\cite{Gegelia}. I thank him for intense and
detailed discussions on that point during this conference, even though no
complete agreement could be reached.

\subsection{Partial Waves in Neutron--Deuteron Scattering}

In the three body sector, the equations to be solved in \ENTNoPion and
ENT(KSW) are computationally trivial and can furthermore be improved
systematically by higher order correction which involve only (partially
analytic, partially numerical) integrations, in contradistinction to
many-dimensional integral equations arising in other approaches. A comparative
study between the theory with explicit, perturbative pions (ENT(KSW)) and the
one with pions integrated out was performed~\cite{pbhg} in the spin quartet
$\mathrm{S}$ wave for momenta of up to $300\;\MeV$ in the centre-of-mass frame
($E_{\mathrm{cm}}\approx70\;\MeV$). As seen above, the two formulations are
identical at LO. Because three body forces enter only at high orders, this
channel is completely determined by two body properties at the first few
orders and no new, free parameters enter.

\begin{figure}[!htb] 
  \centerline{\includegraphics*[width=0.5\textwidth]{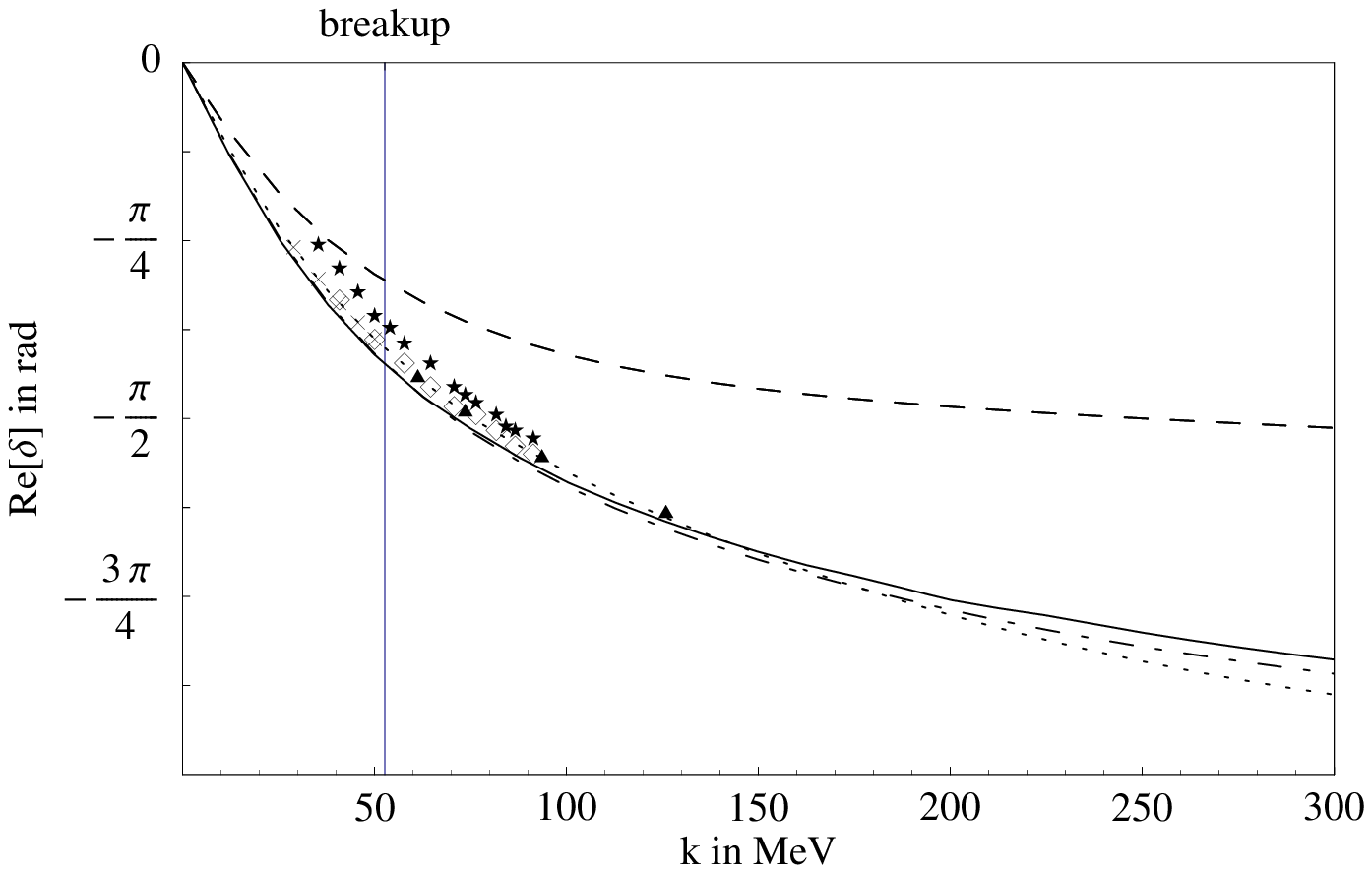}
    \hfill
    \includegraphics*[height=0.415\textwidth,angle=90]{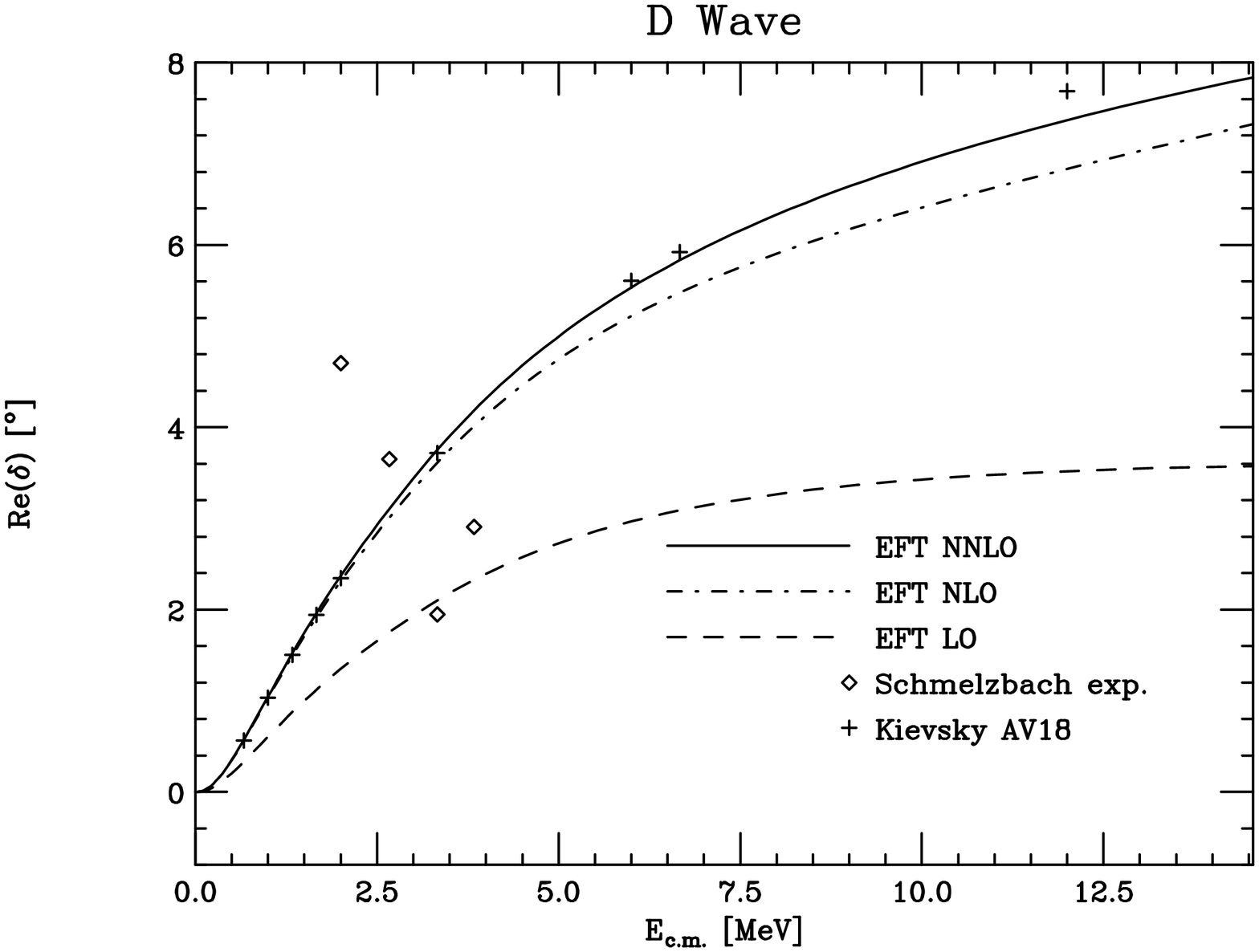}}
\caption{Real parts of the quartet \protect$\mathrm{S}$\protect~\cite{pbhg}
  and doublet \protect$\mathrm{D}$\protect~\cite{pbfghg} wave phase shifts in
  \protect$\mathrm{nd}$ scattering. Legend left: Dashed: LO; solid
  (dot-dashed) line: NLO with perturbative pions (pions integrated out);
  dotted: \NtwoLO without pions. Realistic potential models: squares, crosses,
  triangles.  Stars: $\mathrm{pd}$ phase shift analysis.}
\label{fig:delta}
\end{figure}

The calculation with/without explicit pions to NLO/\NtwoLO shows convergence:
For example, the scattering length is $a(\mathrm{LO})=(5.1\pm 1.5)\;\fm$,
$a(\mathrm{NLO},\NoPion)=(6.7\pm 0.7)\;\fm$, and
$a(\mbox{\NtwoLO},\NoPion)=(6.33\pm 0.1)\;\fm$~\cite{Stooges2}. The
experimental value is $a({}^4\mathrm{S}_\frac{3}{2},\mathrm{exp})=(6.35\pm
0.02)\;\fm$.  Comparing the correction of the scattering length as each order
is added provides one with the familiar error estimate at \NtwoLO:
$(\frac{1}{3})^3\approx 4\%$.  The \NtwoLO calculation is inside the error
ascertained to the NLO calculation. The calculation of pionic corrections in
ENT(KSW) shows that they are -- although formally NLO -- indeed much weaker.
The difference to \ENTNoPion should appear for momenta larger than $\mpi$
because of non-analytic contributions of the pion cut, but those seem to be
very moderate, see Fig.~\ref{fig:delta}. This and the lack of data makes it
difficult to assess whether the KSW power counting scheme to include pions as
perturbative increases the range of validity over the pion-less theory.

Finally, the real and imaginary parts of the higher partial waves
$l=1,\dots,4$ in the spin quartet and doublet channel were found~\cite{pbfghg}
in a parameter-free calculation in \ENTNoPion, see Fig.~\ref{fig:delta}.
Within the range of validity, convergence is good, and the results agree with
potential model calculations (as available) within the theoretical
uncertainty. That makes one optimistic about carrying out higher order
calculations of problematic spin observables like the nucleon-deuteron vector
analysing power $A_y$ where EFT will differ from potential model calculations
due to the inclusion of three-body forces.

\section{Outlook}

Many questions remain open: Which is the the power counting Nature chose in
the pion-ful theory for the coupling of two and more nucleons to external
currents? Does one there include pions perturbatively or non-perturbatively?
At the moment, some people advocate a mixture. To investigate processes with
pions in the initial or final state might be helpful~\cite{bbhg}. How to
extend the analysis of \ENTNoPion in the triton channel systematically to
higher orders?  Technically, how to regularise a Faddeev equation numerically
with external currents coupled in a field theory?  How do four body forces
scale? Extend to nuclear matter!



\begin{thebibliography}{99}
  
\bibitem{review} S.R.~Beane, P.F.~Bedaque, W.C.~Haxton, D.R.~Phillips and
  M.J.~Savage, nucl-th/0008064.  

\bibitem{polarisabilitiesrunning} H.W.~Grie\3hammer and Th.R.~Hemmert, in
  preparation.
  
\bibitem{comptonnopions} G.~Rupak and H.W.~Grie\3hammer, nucl-th/0012096.
  
\bibitem{Compton} J.-W.~Chen, H.W.~Grie\3hammer, M.J.~Savage and R.P.~Springer,
  Nucl.~Phys.~\textbf{A644}, 221 (1998); Nucl.~Phys.~\textbf{A644}, 245
  (1998).
  
\bibitem{pbhg} P.F.~Bedaque and H.W.~Grie\3hammer, Nucl.~Phys.~\textbf{A671},
  357 (2000).
  
\bibitem{pbfghg} F.~Gabbiani, P.F.~Bedaque and H.W.~Grie\3hammer,
  Nucl.~Phys.~\textbf{A675}, 601 (2000).

\bibitem{Epelbaum} E.~Epelbaum, these proceedings.  
  
\bibitem{Phillips} D.R.~Phillips, these proceedings.  
  
\bibitem{Timmermans} R.~Timmermans, these proceedings.  
  
\bibitem{Birse} M.~Birse, these proceedings.  
  
\bibitem{Weinberg} S.~Weinberg, Nucl.~Phys.~\textbf{B363}, 3 (1991).
  
\bibitem{CRS} J.-W.~Chen, G.~Rupak and M.J.~Savage, Nucl.~Phys.~\textbf{A653},
  386 (1999).
  
\bibitem{KSW} D.B.~Kaplan, M.J.~Savage and M.B.~Wise,
  Phys.~Lett.~\textbf{B424}, 390 (1998); Nucl.~Phys.~\textbf{B534}, 329 (1998).

\bibitem{BiraAleph} U.~van Kolck, \journal{\NPA}{645}{273}{1999}.

\bibitem{Beane:2001bc}
S.R.~Beane, P.F.~Bedaque, M.J.~Savage and U.~van Kolck,
nucl-th/0104030.
  
\bibitem{Stooges2} P.F.~Bedaque, H.W.~Hammer and U.~van Kolck,
  Phys.~Rev.~Lett.~\textbf{82}, 463 (1999); Nucl.~Phys.~\textbf{A646}, 444
  (1999); Nucl.~Phys.~\textbf{A676}, 357 (2000).

\bibitem{HammerMehen}
H.W.~Hammer and T.~Mehen,
nucl-th/0105072.

\bibitem{hgtriton} H.W.~Grie\3hammer, in preparation.

\bibitem{Gegelia} J.~Gegelia, these proceedings.

\bibitem{bbhg} B.~Borasoy and H.W.~Grie\3hammer, nucl-th/0105048.
  
\end{thebibliography}
\end{document}